\documentclass[twocolumn]{aa}  
\usepackage{graphicx,color}
\usepackage{natbib} 
\bibpunct{(}{)}{;}{a}{}{,} 
\usepackage{multirow}

\newcommand{\degree}{$^\circ$}

\newcommand{\kms}{km\,s$^{-1}$}
\newcommand{\ulyss}{\emph{ULySS}}
\newcommand{\hi}{H\,{\sc i}}
\newcommand{\hb}{H$\beta$}
\newcommand{\oiii}{[O\,{\sc iii}]}
\newcommand{\nn}{NGC\,404}

\newcommand{\nb}[1]{{#1}}

\usepackage{txfonts}
\begin{document}
   \title{Stellar population and kinematics of NGC404}


   \author{
          A. Bouchard\inst{1,2},
          Ph. Prugniel\inst{2},
          M. Koleva\inst{2,\nb{3,4}}
          \and
          M. Sharina\inst{5\nb{,6}}
          }
\authorrunning{Bouchard et al.}
\titlerunning{Stellar population and kinematics of NGC404}

   \offprints{A. Bouchard}
   \institute{
Department of Astronomy, University of Cape Town, Private Bag X3, Rondebosch 7701,Republic of South Africa
 \and
Universit\'e de Lyon, Lyon, F-69000, France ; Universit\'e Lyon~1,
Villeurbanne, F-69622, France; Centre de Recherche Astrophysique de
Lyon; Observatoire de Lyon, St. Genis Laval, F-69561, France ; CNRS, UMR 5574
 \and
\nb{Instituto de Astrof\'{\i}sica de Canarias, La Laguna, E-38200 Tenerife, Spain
\and
Departamento de Astrof\'{\i}sica, Universidad de La Laguna, E-38205 La Laguna, Tenerife, Spain}
 \and
Special Astrophysical Observatory of the Russian Academy of Sciences, Nizhnij Arkhyz, Karachaevo-Cherkessia, Russia
\and
\nb{Isaac Institute Chile, SAO Branch, Casilla 8-9, Correo 9 , Santiago, Chile.}
}
   \date{\today}

  \abstract
   {NGC404 is a nearly face-on nearby low-luminosity lenticular galaxy. Probing its characteristics provides a wealth of information on the details of possible evolution processes of dS0 galaxies which may not be possible in other, more distant objects.}
   {We study the internal kinematics and the spatial distribution of the star formation history in NGC404. }
   {We obtained long slit spectroscopy at the OHP 1m93 telescope along the major and minor axes of NGC404. The spectra have a resolution $R=3600$ covering a wavelength range from 4600 to 5500\,\AA{}. The data are fitted against the Pegase.HR stellar population models to derive simultaneously the internal stellar kinematics, ages and metallicities. All this is done while taking into account any instrumental \nb{contamination} to the line of sight velocity distribution. Firstly, the global properties of the galaxy are analyzed by fitting a single model and to the data and looking at the kinematic variations and SSP equivalent age and metallicities as a function of radius. Afterwards, the stellar populations are decomposed into 4 components that are individually analyzed.}
   {NGC404 clearly shows two radial velocity inversions along its major axis. The kinematically decoupled core rotates in the same direction as the neutral hydrogen shell that surrounds the galaxy. We resolved the star formation history in the core of the galaxy ino 4 events: A very young ($< 150$\,Myr, and [Fe/H] = 0.4) component with constant on-going star formation, a second young (430\,Myr) component with [Fe/H] = 0.1, an intermediate population (1.7\,Gyr) which has [Fe/H] = -0.05 and, finally, an old (12\,Gyr) component with [Fe/H] = -1.26.  The two young components fade very quickly with radius, leaving only the intermediate and old population at a radius of 25\arcsec{} \nb{(370\,pc)} from the centre.}
   {We conclude that NGC404 had a spiral morphology about 1 Gyr ago and that one or many merger events has triggered a morphological transition. The interstellar medium in the galaxy has two components, the cold molecular gas is most probably a remnant from its past spiral incarnation and the outer neutral hydrogen layer which has probably been acquired in one of the latest merger. }

   \maketitle

\section{Introduction}

Although at the morphological intersection between spirals (S) and ellipticals (E), there may be more to lenticular galaxies (S0) than meets the eye. Indeed, while global properties of more massive S0 galaxies closely resembles those of giant Es, this is not the case for fainter objects \citep[see][]{vandenbergh2009}. Rather than being formed by the successive infall of galactic satellites as seem to be the case for giant Es, the smaller S0s appear to essentially be remnants of gas depleted spirals \citep{bedregal2008}. Several processes, including environmental effects \citep{moran2006} or galactic outflows caused by intense star formation \citep{davidge2008} are possible causes for the ISM exhaustion and thereby the transformation of spirals into S0s. These evolutionary scenarios may be particularly prominent in low-mass galaxies, as their weaker gravitational potential make them more subject to gas depletion.

Recent discoveries of spiral arms, bars and disks in several dEs and dS0s \citep{jerjen2000, barazza2002, graham2003, derijcke2003, derijcke2004, chilingarian2007, lisker2007} have put a new twist in the debate on these objects origin. It was proposed that different formation mechanisms and progenitors were required to account for the diversity of the characteristics of the early-type dwarfs (nucleation, flattening, rotation, ...) \citep{vanzee2004,lisker2008,boselli2008}. Dwarf Irregulars (dIrrs) or BCDs may evolve into dEs and low luminosity S into dS0s \citep{aguerri2005}. 

At $\sim$3.4\,Mpc \citep{tonry2001, karachentsev2002, tikhonov2003}, \object{NGC404} is the closest dS0 galaxy to the Local Group (see Table \ref{prop} for a summary of the targets properties). It is also fairly isolated, the nearest neighbor being over 1\,Mpc away \citep{karachentsev1999}. This makes this nearly face-on galaxy \citep{barbon1982,delrio2004} a good test case for evolution scenarios as one would expect minimal recent impact from any environmental effects \citep[see][]{bouchard2009}. This relative isolation coincides with active star formation \citep[SF,][]{ho1993} and a dust lane with complex structure within 5\arcsec{} from the centre \citep{barbon1982, gallagher1986, wiklind1990, ravindranath2001}. \citet{delrio2004} found a ring of neutral hydrogen (H\,{\sc i}) ($M_{HI} = 1.5\times10^8\,M_{\odot}$) surrounding the stellar disk, indicating that internal processes (most likely stellar feedback from intense SF within the molecular clouds) are evacuating the H\,{\sc i} gas from the centre of the galaxy. 
NGC404 has also a low-ionization nuclear emission line region \citep[LINER, ][]{pogge2000}.

By comparison, NGC404 is only about 0.8 magnitude brighter than NGC205 or the other dEs in nearby clusters presented in \citet{koleva2009b}. Should the remaining gas in this galaxy be expelled, stripped or otherwise exhausted, the stellar population would fade and any remaining morphological `asymmetries' would vanish: this object would become very similar to any other early-type dwarf. In this paper we examine this possibility from the view point of the stellar populations. Using long-slit optical spectroscopy along the major and minor axis (Figure \ref{image}), we measure radial variations of the star formation (SF) and metal enrichment histories within the central arc-minute of NGC404. 

This paper is organized as follows: Section 2 presents the data acquisition, reduction and analysis strategy used in this paper. Section 3 presents age, metallicity and kinematical distribution profiles for this galaxy. In Section 4 we discuss a more in-depth approach to the SF history analysis. The overall discussion is found in Section 6 while the main conclusions are recalled in Section 7.

\begin{table}[tb]
\begin{center}
\caption{Properties of NGC404}\label{prop}
\begin{tabular}{l c c}
\hline
\hline
Parameter & Value & Ref.\\
\hline
R.A. (J2000)				& 01$^h$09$^m$27.0$^s$ 		& \\
Dec. (J2000)				& +35\degree{}43\arcmin{}04\arcsec{} & \\
Type 					& dS0						& \\
$\epsilon$\,=\,$1 - (b/a)$		& 0.07						& 1\\
\nb{$D$ (Mpc)}				& \nb{3.05$\pm$0.042}				& \nb{2}\\
$M_B$					& $-16.26\pm$0.15 				& 3\\
$M_V$					& $-17.63\pm$0.15 				& 1\\
$D_{25}$					& 3.5\arcmin{} 					& 4\\
						& \nb{3.1\,kpc}\\
 
$r_e$ (\emph{bulge}) 		& 61\arcsec{} 					& 5\\
						& \nb{0.9\,kpc}\\
$M_{\rm HI}$				& 1.5$\times10^8$\,$M_{\odot}$ 	& 6\\
$M_{\rm HI}/L_B$			& 0.22 $M_{\odot}$/$L_{\odot}$ 	& 6\\
\hline
\end{tabular}
\end{center}
References. (1)~\citet{tikhonov2003}; \nb{(2)~\citet{dalcanton2009}; }(3)~\citet{karachentsev2002}; (4)~\citet[RC3,][]{devaucouleurs1991}; (5)~\citet{baggett1998}; (6)~\citet{delrio2004}
\end{table}

\section{Observations}

\begin{figure}
\includegraphics[width=0.45\textwidth]{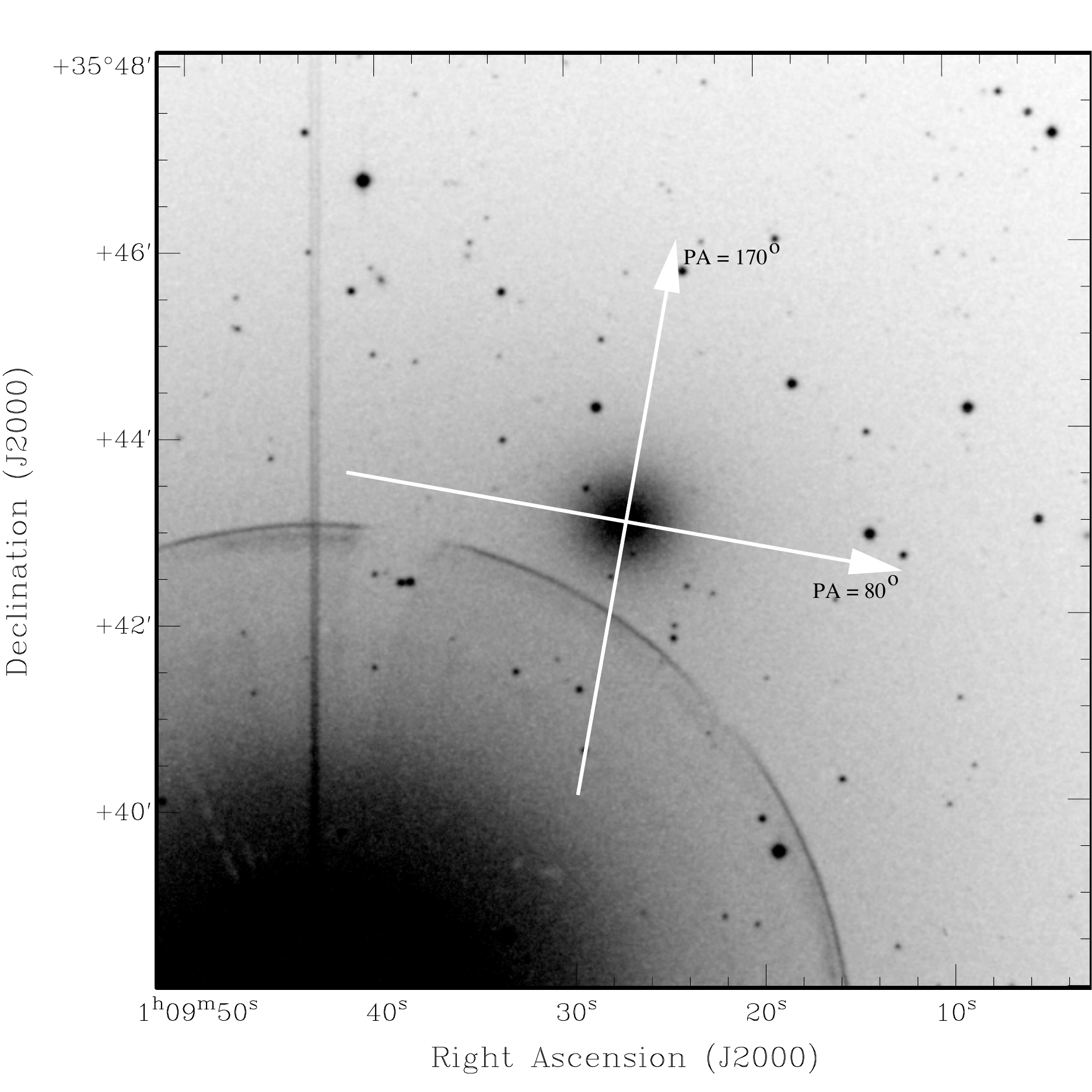}
\caption{DSS-2 image of \nn{}. The two observed PAs are shown in white, the arrows indicate the direction of the spatial axis (see Figure~\ref{rad_prof}). The second magnitude foreground star $\beta$\,And contaminates the South-East corner of the image.}\label{image}
\end{figure}

We used the OHP 1m93 telescope in November and December 2005 to obtain long-slit spectroscopic data of NGC404 along 2 position angles (PA = 80$^\circ$ and PA = 170$^\circ$, also referred to as the major and minor axis respectively,  see Figure\,\ref{image}). The CARELEC spectrograph with the 1200 grating yielded a nominal dispersion of 33\AA{}/mm \nb{and a wavelength coverage from 4600 to 5500 \AA{}}. It has a central wavelength of $\lambda_c$ = 5108~\AA{} and $\lambda/\Delta \lambda$ = 3600 (FWHM; instrumental velocity dispersion $\sigma_{ins}$\,=\,35 \kms{})\footnote{Measured on twilight spectra, see Fig.~\ref{lsf}. The resolution relative to the Elodie library is $\sigma_{ins}$\,=\,33 \kms{}.} and was equipped with a 0.325 mm slit. The EEV CCD (2048$\times$1024 pixels) had pixel of 13.5\,$\mu$m or 0.45\,\AA{}$\times$0.54$^{\prime\prime}$. A total of 6 one-hour frames were taken for the PA\,=\,80$^\circ$ slit position and 7 one-hour frame for PA\,=\,170$^\circ$. 

The galaxy has a very bright foreground star (\object{$\beta$\,And}, $m_V = 2$) approximately 5\arcmin{} to the South-East. The star is contaminating the spectra for radii greater then $\sim$30 arcsec from the centre of the galaxy.

\subsection{Data Reduction}
\label{reduction}
The data were reduced using the IRAF longslit spectroscopy package. After the bias subtraction, flat fielding, wavelength calibration and sky subtraction, the multiple exposures were co-added to produce one final image for each PA. No special effort has been made to get rid of cosmic rays or other CCD defect as these can be handled later during the spectral fitting stage; there is a moderate number of such defects. 

For the radial profile analysis (see Section \ref{radial}), the spectra were boxcar smoothed along the spatial axis to increase their signal to noise (S/N) ratio. In order to avoid smearing out the information in the central region, the smoothing kernel was taken to be 3 spatial pixels ($\sim1.5\arcsec$ \nb{or 22\,pc at the distance of NGC404}) at the centre of the galaxy and increased linearly with radius to 29 pixels ($\sim15\arcsec$ \nb{or 220\,pc}) at angular offsets of $\pm60\arcsec{}$. This resulted in S/N$\sim$50 at $R$\,=\,0\arcsec{} and all spatial positions where S/N $\ge$ 1 were analyzed. 

For the more detailed stellar population analysis presented in Sections \ref{stellarpop} and \ref{sfh_section} the spectrum for the central region is an average of all spectra within 5\arcsec{} \nb{(75\,pc)} from the centre of the galaxy (both PAs).  The outer spectrum (at $R$\,=\,25\arcsec{} \nb{or 370\,pc}) is similarly a combination of all spectra between 15\arcsec{} and 35\arcsec{} from the centre of NGC404.

\subsection{Full Spectrum Fitting}

The analysis was performed with \ulyss{}\footnote{\ulyss{} is available at http://ulyss.univ-lyon1.fr} \citep[Universit\'e de Lyon Spectroscopic analysis Software,][]{koleva2009a}, a full spectrum fitting software package. It fits an observed spectrum against a linear combination of non-linear model components, convolved with a line-of-sight velocity distribution (LOSVD) and multiplied by a polynomial to absorb the effects of the flux calibration uncertainties and of the internal or Galactic extinction \nb{(see section \ref{dust} for more details)}. The advantages of this program are (i) to use all the pixels (weighted by their inverse squared estimated errors) and (ii) to minimize simultaneously all the parameters leading to the optimal solution despite the degeneracies. This program is used to fit single age and single metallicity population (SSP) models or composite models including the nebular emission lines and one or several SSPs.

\begin{figure*}
\begin{center}
\includegraphics[width=0.95\textwidth]{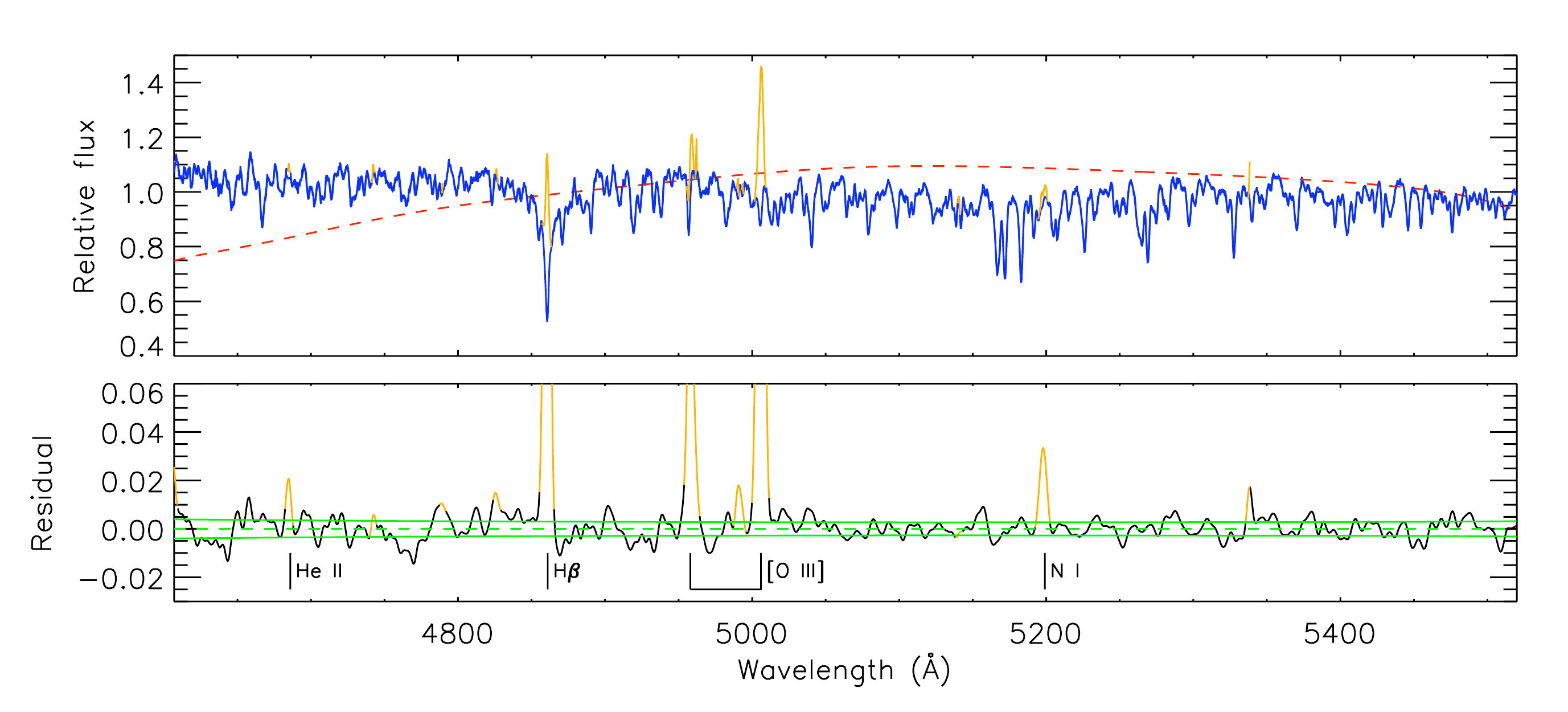}
\caption{Example fit as performed with \ulyss{}. The top panel shows the data (black), the best fitting model (blue) and the normalising polynomial (dashed red). The parts of the spectra that were ignored in the fitting procedure are plotted in orange. The bottom panel shows the fit residual smoothed to 100\,km\,s$^{-1}$ resolution; the green lines are the mean residual value (dashed) and $\pm1\sigma$ residual deviation (solid). The best fitting model here is a 4 component model as described in section \ref{sfh_section}.}\label{example_fit}
\end{center}
\end{figure*}

The SSP spectra were generated using Pegase.HR \citep{leborgne2004} and the ELODIE 3.1 stellar library \citep{prugniel2001, prugniel2007} using \citet{salpeter1955} initial mass function. A fit example is provided in Figure \ref{example_fit}; the actual fitted model is a 4 population composite model and is described in Section \ref{sfh_section}.

\begin{figure}
\includegraphics[width=0.45\textwidth]{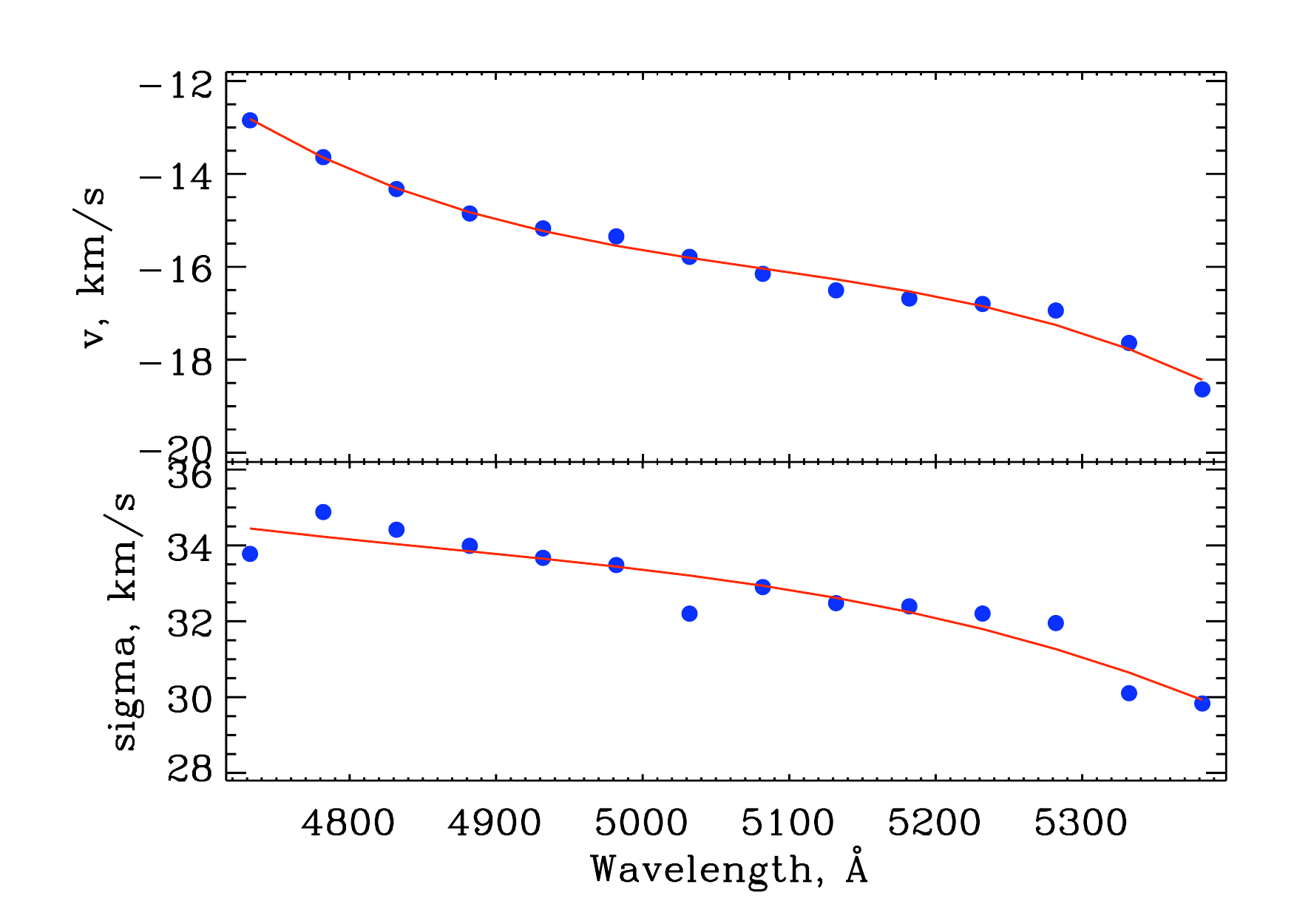}
\caption{Relative LSF ($v_{rad}$, top panel and $\sigma_{ins}$, bottom panel) between the observed spectra and the model as a function of wavelength. The blue dots represent the measured LSF, while the solid line is the smoothed version used as input for the rest of the analysis.}\label{lsf}
\end{figure}

The first step in the \ulyss{} analysis is to evaluate the relative line spread function (LSF) between the models and the observed CARELEC spectra. This was done by comparing twilight sky spectra with a solar spectrum. The relative gaussian broadening and velocity offset was determined with {\sc uly\_lsf} in overlapping windows of 250\,\AA{} separated by 50\,\AA{}. The results are shown in Figure \ref{lsf}. The velocity residuals, changing from $-13$ to $-19$\,\kms, come from the uncertainty on the wavelength calibration (0.05 \AA{} peak-to-peak). The decrease of the relative instrumental velocity dispersion, $\sigma_{ins}$, from 35 \kms{} (blue) to 30 \kms{} (red) is a characteristic of the spectrograph and grating.

This measured instrumental LSF is in turn used as an input parameter to \ulyss{} which {\it injects} it in the model before performing the minimization (see \citealp{koleva2009a} for details).

\subsection{\nb{Effects of Dust Extinction}}
\label{dust}
\nb{In the case of a foreground dust screen (e.g.\, Galactic extinction), the absorption is a slowly varying function of wavelength and, in that respect, is exactly modeled by the low resolution Legendre polynomial normalization used in \ulyss{} (provided that a single SSP is fitted). 
However, the dust has an inhomogeneous distribution within the galaxy \citep[c.f.,][]{wiklind1990} and the determination of the SFH (determined by fitting multiple SSPs) may be affected if the different generations of stars are differently absorbed.} 

\begin{figure*}
\begin{center}
\includegraphics[width=0.45\textwidth]{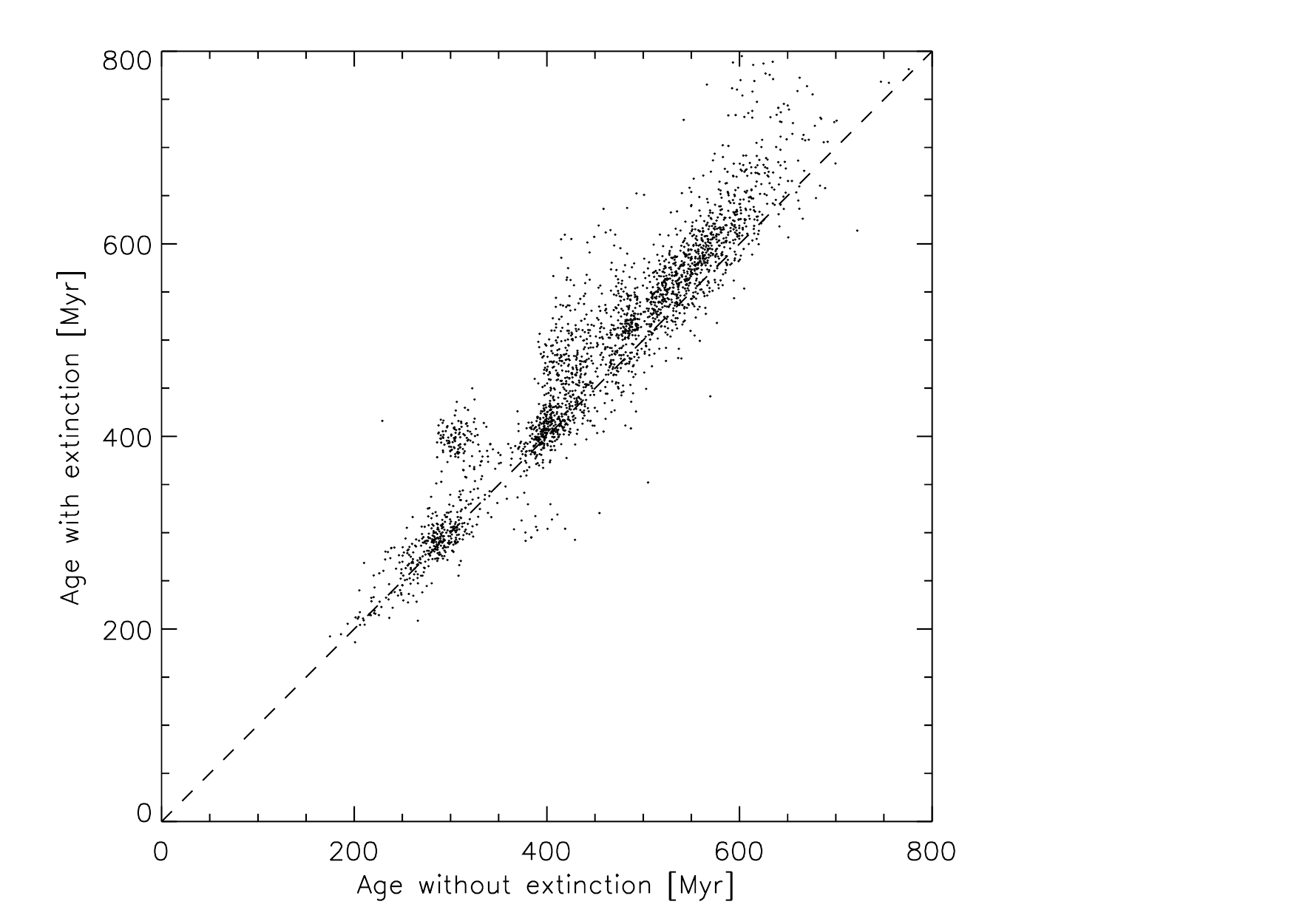}
\includegraphics[width=0.45\textwidth]{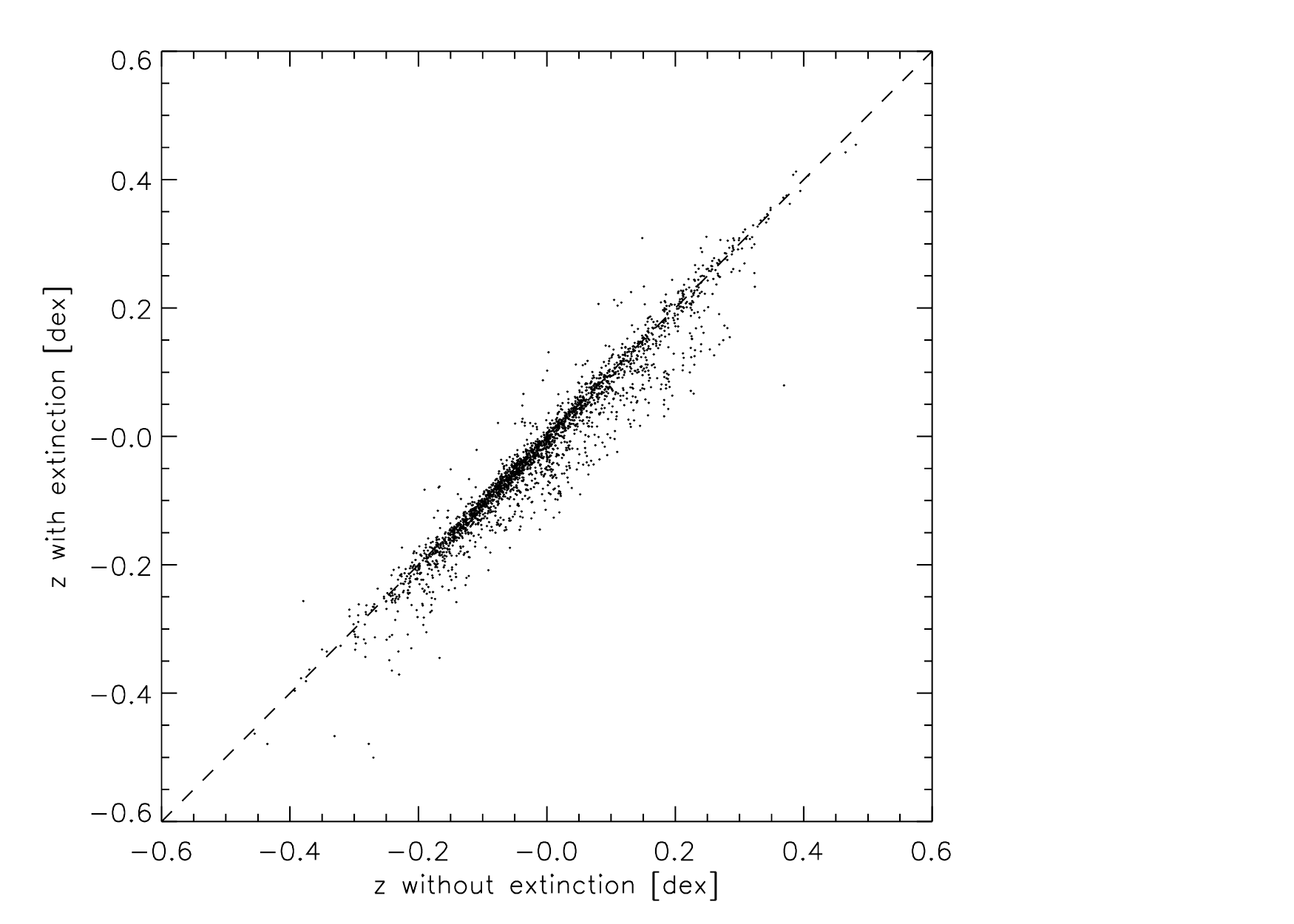}
\caption{\nb{Monte Carlo simulations of the effects of dust on the age (left) and z( right) measure for the youngest SSP component of a 4 component fit.}}\label{dustfig}
\end{center}
\end{figure*}

\nb{Figure \ref{dustfig} compares the results of 2500 Monte-Carlo iterations where two different sets of SSP models (with or without dust)  are used to determine the age of a composite stellar population. In this case, four simultaneous components (fully described in section \ref{sfh_section}) are fitted to the core spectra of NGC404. The value on the x-axis shows the age of the youngest SSP component found when neglecting dust effects compared with the age found when a free amount of dust absorption is added to each component. We adopt the \citet{tojeiro2009} approach and take this effect into account by inserting a specific extinction on young stellar spectra ($<$1\,Gyr). }

\nb{It is clear from Figure \ref{dustfig} that including absorption has marginal effect: neglecting dust makes stars look younger and more metal rich (thereby more appealing) but the effect can be safely ignored has it is smaller than the accuracy of the method (systematic shifts of the order of 50\,Myr and 0.01 dex). The test recovers a very large range in absorption measures with $A_B$\,=\,0.5$\pm$0.4\,mags, which implies that the method is globally insensitive to internal dust absorption: it is doubtful that this value be reliable. In the remainder of the current paper, we neglect any dust component in order to minimize the number of free parameters in the fits.}

%


\section{Radial Profiles}
\label{radial}

Figure \ref{rad_prof} presents luminosity, kinematical ($v_{rad}$ and $\sigma_v$) and SSP equivalent age and metallicity profiles for the stellar population as a function of the radius for the two observed PAs. First the stellar population was adjusted with a single SSP using the automatic clipping of the outlier to reject the emission lines and the defects (option {\sc /clean}). The emission lines were then measured on the fit residuals. Since a preliminary analysis had shown that their kinematics were different, the \oiii{} doublet ($\lambda\,=$\,4958\,\AA{} and 5006\,\AA) and the \hb{} line ($\lambda$\,=\,4861\,\AA) were fitted separately, i.e., independently determining the kinematics of both systems.

\subsection{Stellar and ISM Kinematics}

The $v_{rad}$ profile along the major axis (PA = 80\degree{}) shows a complicated structure with what appears to be a double inversion of the velocity gradient. The core, within $\pm1$\arcsec{} from the centre, is in counter-rotation with a 2.2\,\kms{} amplitude. The surrounding region displays a linear velocity gradient until $\pm20$\arcsec{} \nb{($\pm295$\,pc at the distance of NGC404)} from the center where the apparent rotation reach $\sim23$\,\kms{}. From this radius outward, the rotation velocity decreases to inverse its direction at $\sim35$\,\arcsec{} \nb{(520\,pc)}. At very large radii, the velocity profile is consistent with the results of \citet{delrio2004} who measured an H\,{\sc i} rotational (i.e., deprojected) velocity to be $\sim$200\,\kms. 
Otherwise, no rotation is seen along the minor axis. 

\begin{figure*}[t]
\begin{center}
\includegraphics[width=0.45\textwidth]{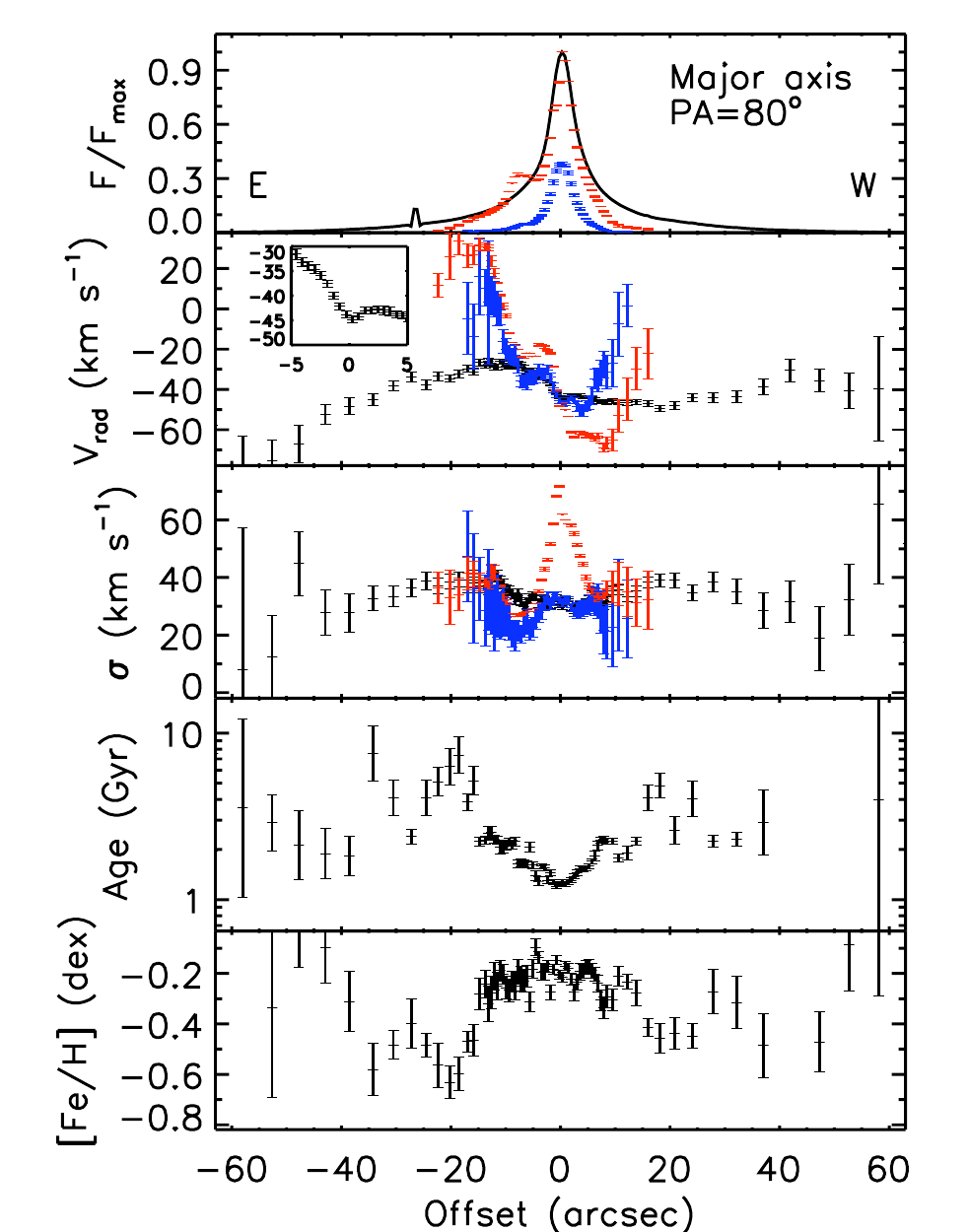}
\includegraphics[width=0.45\textwidth]{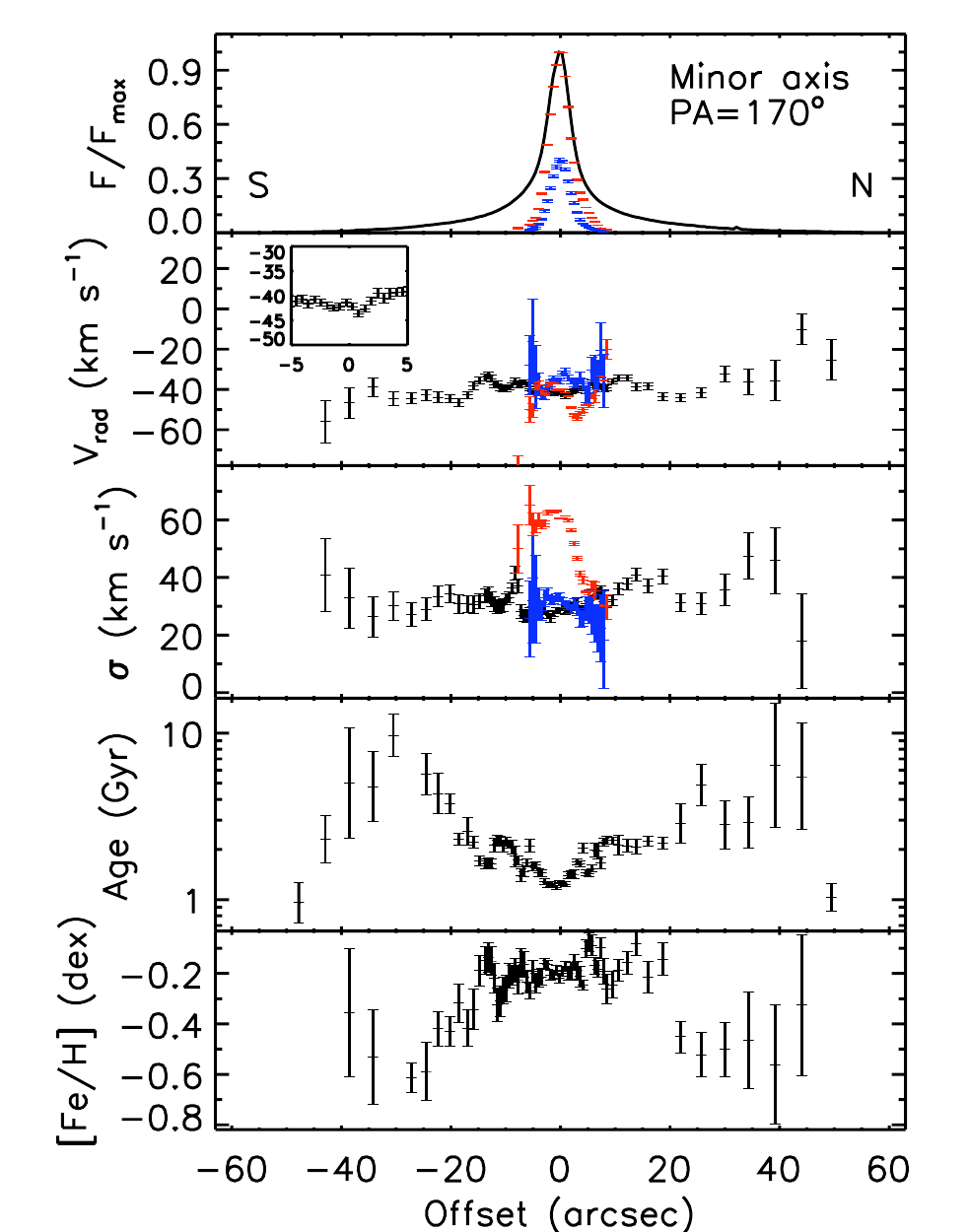}
\caption{Radial profiles for normalised flux (top panel), $v_{rad}$ (2$^{nd}$ panel), $\sigma_v$ (middle panel), Age (4$^{th}$ panel) and [Fe/H] (bottom panel) for the two observed PA of NGC404 (major axis, PA = 80\degree{} on the left and minor axis, PA = 170\degree{} on the right). Overplotted on the top three panels are the profiles derived from the \hb{} (blue) and \oiii{} (red) emission lines. The flux values for \oiii{} are the sum of the $\lambda\,=$\,4958 and 5006\,\AA{} lines and are normalized to the central total flux with \hb{} flux plotted on the same scale. The insets on the second panels shows a zoom on the central stellar velocity profiles.
}\label{rad_prof}
\end{center}
\end{figure*}

The central velocity dispersion, i.e., the average $\sigma_v$ value in the inner 5\arcsec{} \nb{(75\,pc)} for both PAs, is $\sigma_v^c$\,=\,$30\pm2$\,km\,s$^{-1}$. This is substantially lower than previous measurements; \citet{barth2002} had found $\sigma_v = 40\pm3$\,km\,s$^{-1}$ from the near-infrared [Ca\,{\sc ii}] triplet.  Our new value is however comparable to the central velocity dispersion of the slightly fainter NGC205 \citep[$\sigma_v^c$\,=\,20$\pm$1\,\kms{},][]{simien2002}. With $L_B = 10^{8.7}$\,$L_{\odot}$ \citep{karachentsev2002} it is also consistent with the \citet{faber1976} relation for dwarf galaxies \citep[see][]{derijcke2005}.

Spectra near the centre of NGC404 clearly shows \hb{} ($\lambda\,=$\,4861\,\AA) and \oiii{} ($\lambda\,=$\,4958 and 5006\,\AA) emission lines originating from the star forming regions in the ISM and the active nucleus. In Figure \ref{rad_prof} is also shown the flux, radial velocity and velocity dispersion for \hb{} in blue and \oiii{} in red. The intensity profiles of \hb{} along both axes are symmetrical, unlike those of the \oiii{} lines which present a secondary peak around 10\arcsec{} \nb{(150\,pc)} to the East of the center, along the major axis. This off-centered peak corresponds the position of a giant molecular cloud that was detected with CO observations \citep{wiklind1990}. 

Globally, the ionised gas distribution is more flattened than the stellar population: The full widths at 20\% of maximum (FW20) for \hb{} are 8.8\arcsec{} and 6.8\arcsec{} for the major and minor axis respectively and 15.0 and 7.5 \arcsec{} for \oiii{} (excluding the secondary peak). This indicates ellipticities of $\epsilon_{H\beta}$\,=\,0.22 and $\epsilon_{\oiii{}}$\,=\,0.5, compared with $\epsilon_{stars}$\,=\,0.12. \citet{tikhonov2003} have found a stellar ellipticity of 0.07.

The ISM and stellar velocity profiles agrees well within the measured range and the \hb{} width is comparable the stellar velocity dispersion. The \oiii{} emission line, however, has a significantly higher dispersion than \hb{} in the central 10 \arcsec{} along both axes: the central velocity dispersion reaches 70\,\kms. 

\subsection{Stellar Populations}
\label{stellarpop}

The SSP-equivalent age and [Fe/H] profiles (Figure \ref{rad_prof}) show similar behavior for both axes. In the star forming region, i.e., within a radius of 20\,\arcsec \nb{(295\,pc)}, the SSP-equivalent age increases from the central value of 1.3\,Gyr to 1.7\,Gyr while de  metallicity remains relatively constant at $-0.2$. Outside of this region and to the edge of the measurements at about 1 $R_e$, the age raise to 10\,Gyr and the metallicity drops down to $-0.6$. These gradients are qualitatively and quantitatively similar to those observed in other dEs and dS0s \citep{koleva2009b,chilingarian2009}.

\begin{figure}
\includegraphics[width=0.5\textwidth]{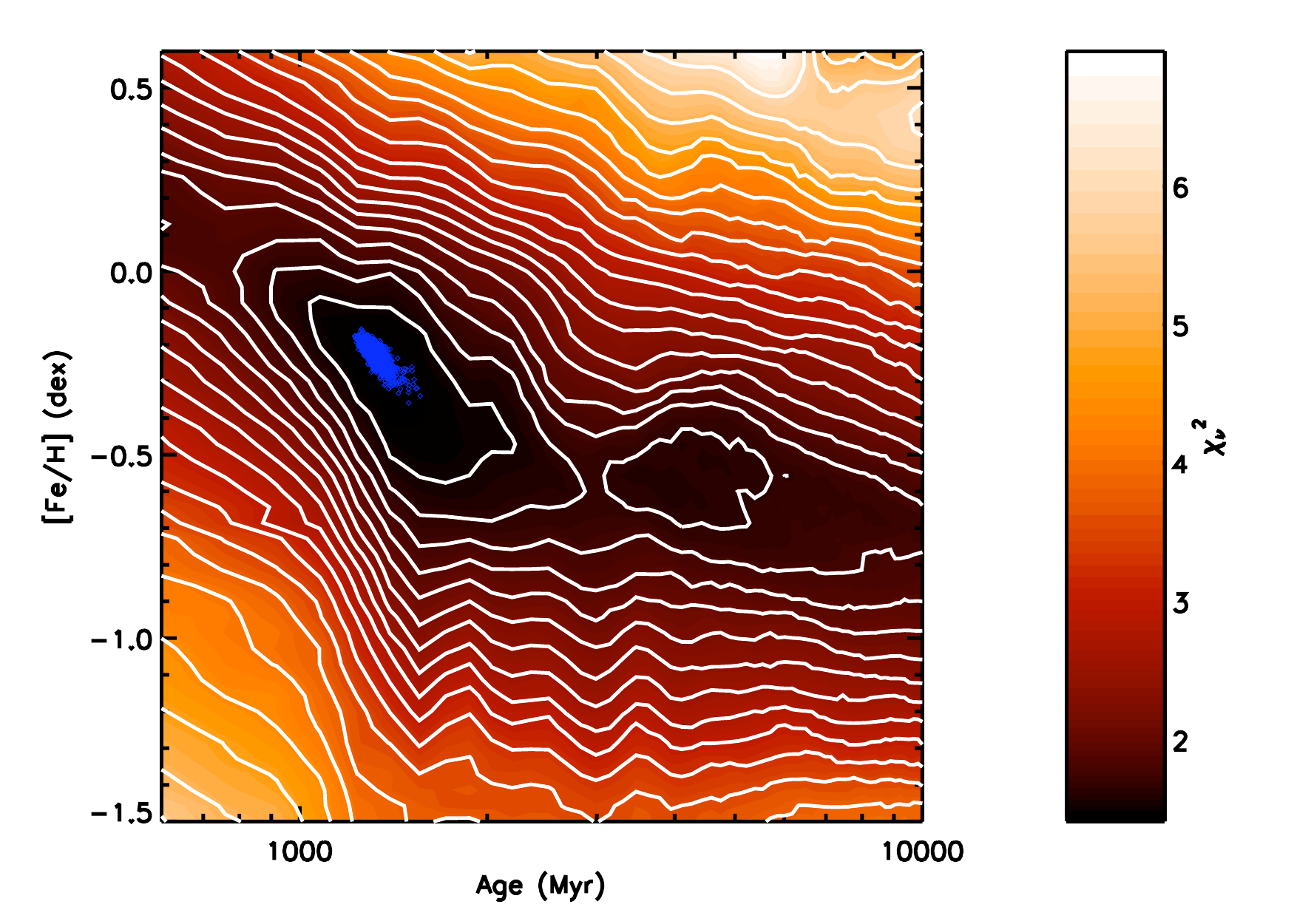}
\caption{$\chi^2_\nu$ distribution map for a single stellar population model fit of NGC404 as a function of age and [Fe/H]. Lines of constant $\chi^2_\nu$ are overplotted (white). The blue dots represent the result of Monte Carlo simulation.}\label{chimap}
\end{figure}

Figure \ref{chimap} shows a  $\chi^2$ map (made with the program {\sc uly\_chimap} in the age vs. metallicity plane) for a spectrum extracted within a radius of 5\arcsec{} \nb{(75\,pc)} from the centre. The map reveals the presence of a local minimum (near 4\,Gyr and $-0.5$\,dex) that requires to perform a global minimization, achieved by using a grid of initial guesses. A series of 2000 Monte-Carlo simulations (blue dots in Fig.~\ref{chimap}), adding each time a  random noise equivalent to the observation noise estimated from the characteristics  of the detector, was used to properly estimate the precision of the measurements. The resulting SSP-equivalent parameters are reported in Table~\ref{centralpop} (age\,=\,$1319\pm41$\,Myr and [Fe/H]\,=\,$-0.225\pm0.016$\,dex).

\begin{table}[tb]
\begin{center}
\caption{Star Formation History}\label{centralpop}
\begin{tabular}{l c c c c}
\hline
\hline
SFH type	& $f_L$ 	& $f_M$ 	& Age 	& [Fe/H] \\
		& (\%)	& (\%)	&(Myr)	& (dex)\\
\hline
\multicolumn{5}{c}{\emph{Core (SSP-equivalent)}}\\
\hline
SSP		& ...				& ...				& 1319$\pm$41	& $-0.225\pm0.016$ \\
\hline
\multicolumn{5}{c}{\emph{Core (4 components)}}\\
\hline
CST		& 4.1$\pm$0.5		& 0.20$\pm$0.03	& 150			& 0.4$\pm0.3$ \\
SSP 		& 20$\pm$1		& 5.6$\pm$0.3		& 430$\pm$100	& 0.1$\pm$0.1 \\
SSP		& 62$\pm$1		& 58$\pm$1		& 1700$\pm$160	& $-0.05\pm0.05$ \\	
SSP		& 14$\pm$1		& 36$\pm$2		& 12000			& $-1.26$ \\
\hline
\multicolumn{5}{c}{\emph{R=25\arcsec{}}}\\
\hline
SSP		& 47$\pm$2		& 37$\pm$1		& 2600$\pm$400	& 0.16$\pm$0.15\\
SSP		& 53$\pm$2 		& 63$\pm$2		& 12000			& $-1.26\pm$0.22 \\
\hline
\end{tabular}
\end{center}
\end{table}

\section{Star Formation History}
\label{sfh_section}
As the presence of extended nebular emission lines in the spectra is evidence for ongoing star formation in the central region of NGC404, it is clear that an SSP is not a good representation of the real stellar population. The radial increase of the SSP-equivalent age (Figure \ref{rad_prof}), may in fact reflect the decreasing fractional contribution of a young and centrally concentrated star formation burst with respect to an older, more uniform stellar population. We therefore reconstruct the star formation history (SFH) of NGC404 following the approach  adopted in \citet{koleva2009b}, i.e., by decomposing the observed spectra in a sum of several components.

We adopted a 4-component model consisting of three SSPs representing various stellar populations in the galaxy, plus one young population with constant, ongoing star formation rate (CST). The latter component was also produced using Pegase.HR with the Elodie~3.1 library and Salpeter IMF.

The age limits between each component were \nb{essentially chosen on a trial and error basis. The method consists in choosing an initial set of populations with the age limits distributed evenly on a logarithmic axis. If any of the fitted populations returns a result that is equal to its boundary limit then that `population box' is either merged with the adjacent one or the limits are modified. In some cases, mainly when a population shows incoherent behavior after a box merger or has non-reproducible results under a global minimization attempt (i.e. providing multiple initial guesses inside the set limits), then an adjacent box is split in two to increase resolution and the process carries on. The solution is considered valid when none of the population's parameter reach any of the set limits and is robust under a Monte Carlo style approach.}

In the end, the limits were set as follows:
\begin{itemize}
\item the old SSP as a fixed age\,=\,12\,Gyr
\item the intermediate age SSP has  800\,$<$\,age\,$<$\,6000\,Myr 
\item the youngest SSP has age\,$<$\,800\,Myr 
\item the CST population began its star formation 150\,Myr ago
\end{itemize} 
No further limits were imposed on metallicity.

This 4-component analysis was applied at two different locations: first at the edge of the star forming region ($R=25$\arcsec{} \nb{or 370\,pc}) and then in the core of NGC404 ($R=0$\arcsec{}, see Section~\ref{reduction} for more details about these spectra). The results are summarized in Table~\ref{centralpop}: the first column shows the type of stellar population that was fitted, the second and third column are the relative stellar luminosity and stellar mass fraction for each of the component, column~4 shows the measured age of each population and finally the last column contains their metallicity. \nb{The results are more significant than the single SSP fit and the $\chi^2_\nu$ is reduced from 1.4 (single SSP) to 1.001 (4 component). }


\nb{Old, low metallicity ([Fe/H]\,=\,$-1.26$ dex) stars contribute to 63\% of the stellar mass just outside of the star formation region, taking into account stellar mass loss due to normal stellar evolution.
There is also a large fraction (37\%) of intermediate age (2.6\,Gyr) and high metallicity (0.16 dex) stars. As expected, we  found no traces of any young population (SSP or CST); no emission lines are seen in the spectra at that location. Our results for the old population are in good agreement with the those from HST stellar photometry ([Fe/H]\,=\,$-1.11$\,dex, \citealp{tikhonov2003}).}

Similar analysis was done in the core (R\,=\,0\arcsec{}). Although old stellar populations of early type dwarfs often exhibit radial metallicity gradients \nb{\citep{harbeck2001,koleva2009b}}, we kept the metallicity fixed at [Fe/H]\,=\,$-1.26$ (the value found at $R=25$\arcsec{}). This was done partly to minimize the number of free parameters in the fit and because, in the presence of a strong young population (luminosity-wise), there is a limit to the amount of information on the older stars that can be extracted from a given spectrum. 

As expected from Figure \ref{rad_prof}, the mass fraction contributed by old stars decreased to 36\%, giving way to a more important (58\%) and somewhat younger (1.7 Gyr) intermediate population with solar metallicity ([Fe/H] = $-0.05$). We also see a non-negligeable (6\%) young stellar population (430\,Myr) and traces of on-going star formation within the last 150\,Myr (0.2\% in mass).

\begin{figure}[bt]
\includegraphics[width=0.45\textwidth]{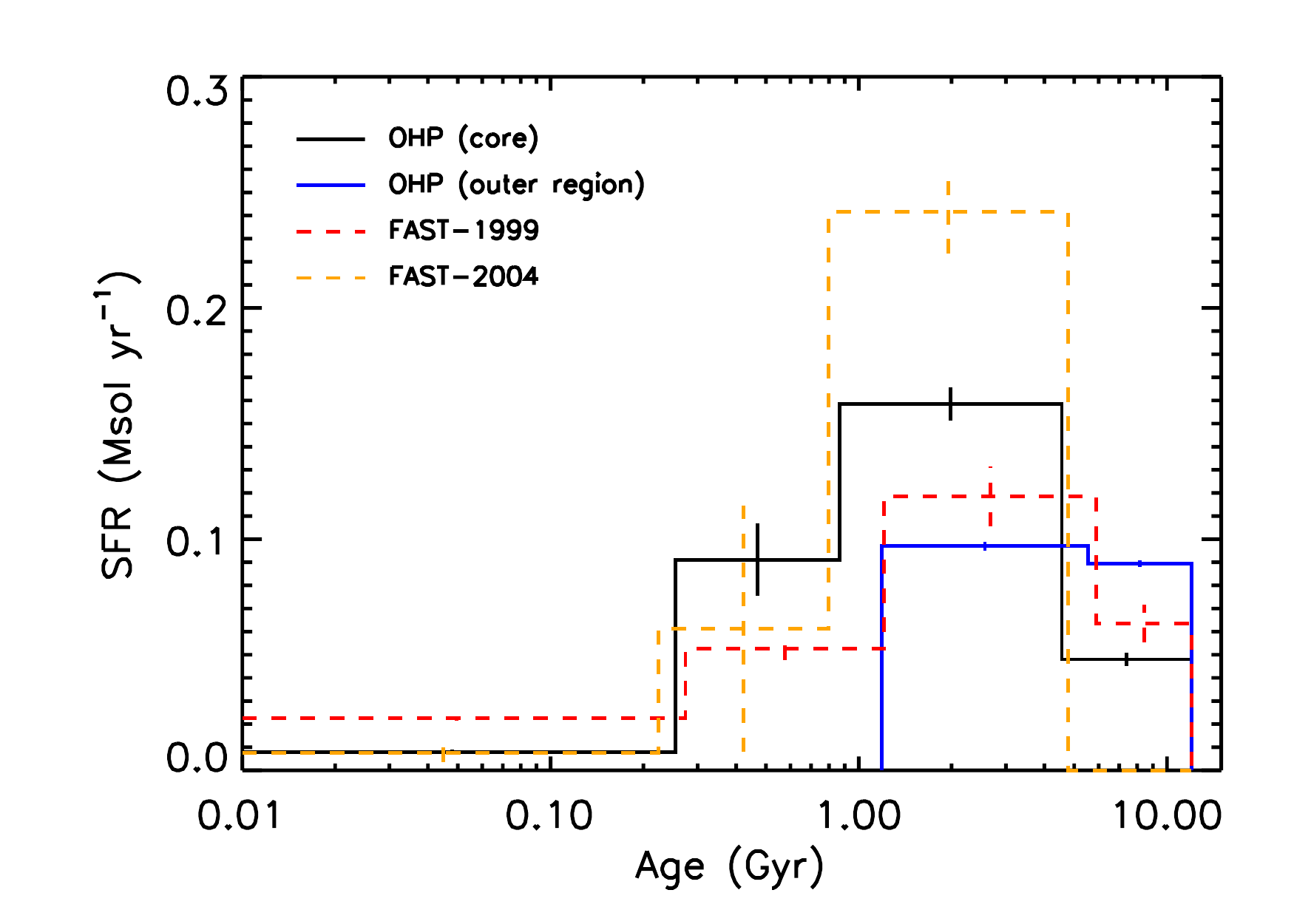}
\caption{SFR has a function of lookback time for an assumed total stellar mass of 1$\times10^9$\,M$_{\odot}$. The black line shows the SFR from the centre of NGC404, the blue shows the outer region. For comparison, we have also plotted results of two FAST spectra, the 1999 data in red and 2004 in gold.}\label{sfh_box}
\end{figure}

To verify the reproducibility and accuracy of our results in the centre of NGC404, we have used two additional spectra from the Fast Spectrograph for the Tillinghast Telescope \citep[FAST,][]{fabricant1998} archive\footnote{http://tdc-www.harvard.edu/instruments/fast/}. The data are 120\,sec exposures on a 1.5m telescope with 1.47\,\AA\,pixel$^{-1}$, covering from 3700 to 7500\,\AA{}. The spectra were obtained on taken on 12-Oct 1999 and 19-Jul 2004.

Figure \ref{sfh_box} presents the results of a similar, 4-component model analysis, on all four spectra. It shows the required SFR as a function of age (lookback time) to produce the measured stellar population ages and relative importance. \nb{The age limits for the bins are chosen the be the halfway point between the measured age of the components, also imposing a maximum age of 12\,Gyr. Although they have more or less comparable overall masses (Table \ref{centralpop}), the SFR associated with the old stellar population is lower but distributed on a longer period than that of the intermediate population.}

The results for the 3 core spectra are consistent on a qualitative level: all 3 have some level of on-going star formation and similar distribution for the three SSP component. The only exception is for the FAST-2004 spectra which shows no evidence for the old stellar population but a stronger intermediate age one. The outer region seems to have had a more uniform SFH, indicating that the star formation burst has been concentrated in the inner few arc-seconds. 

\begin{figure}
\includegraphics[width=0.45\textwidth]{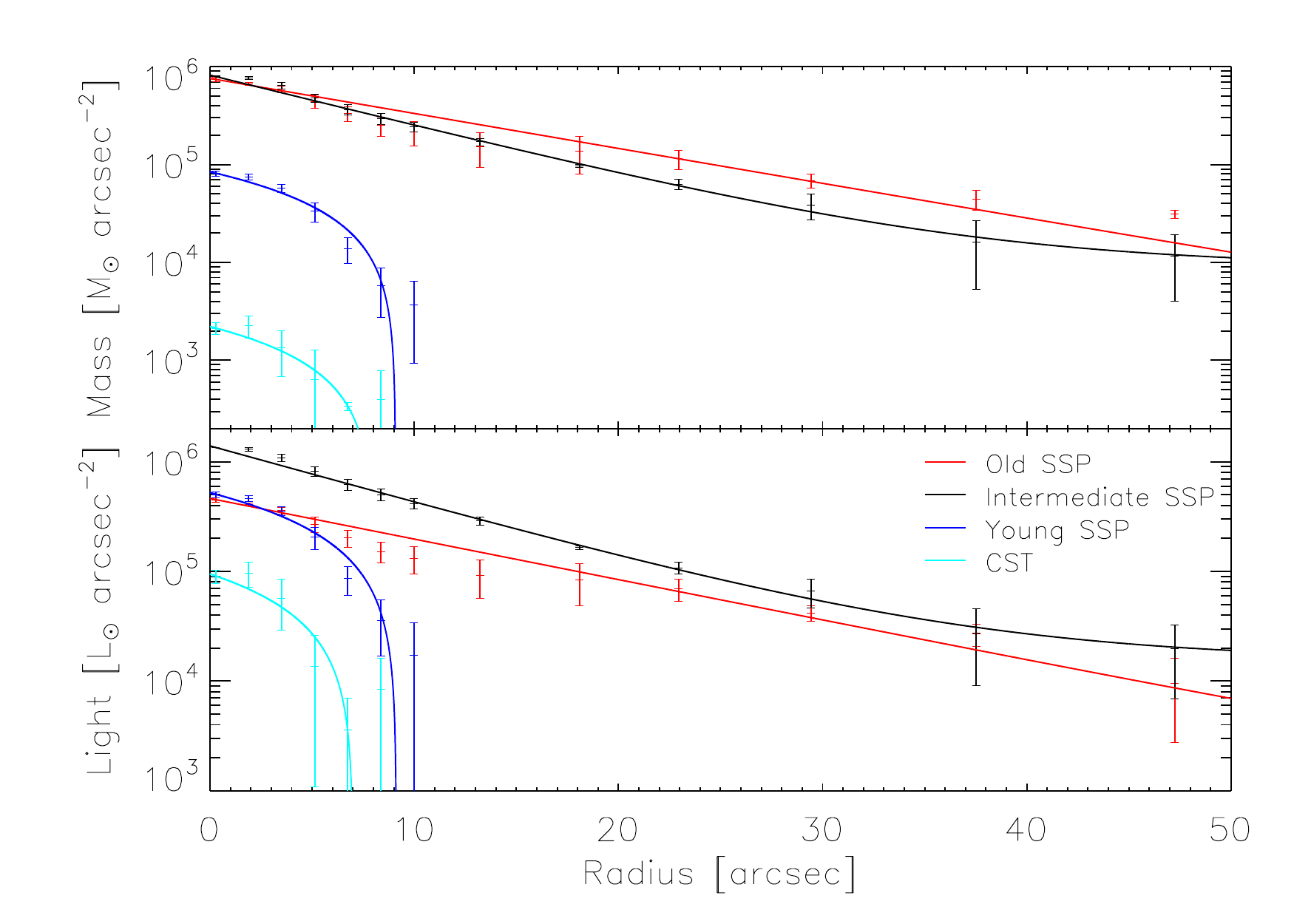}
\caption{Stellar mass (\emph{top}) and stellar light (\emph{bottom}) distribution of each population as a function of radius. The old stellar population is in red, the intermediary in black, the young in dark blue and the very young population with constant SFR is in cyan. \nb{The lines show the results of exponential fits through the data points.}}\label{sfh_rad}
\end{figure}

We have also traced the stellar light and mass distribution for each of the 4 components as a function of radius as shown in Figure~\ref{sfh_rad}. To accomplish this, the age and metallicity of each component were fixed to the values found in the core of NGC404 (Table~\ref{centralpop}), leaving only the relative fraction to vary with radius. These mass and light fractions resulting from the various fits were then converted to an absolute distribution by scaling our results to the values measured by \citet{tikhonov2003} \nb{and \citet{dalcanton2009}, i.e.,  $\mu_{V,0}$\,=\,16.6, $D$\,=\,3.05\,Mpc respectively and the stellar mass to light ratio derived from the fitted models (on average $M/L_V$\,=\,0.7).}

We can indeed see from Figure \ref{sfh_rad} that the two young components are both centrally concentrated and have similar distribution: they are found within 10\arcsec{} \nb{(150\,pc)} from the core of NGC404. This corresponds to the size of the molecular cloud found by \citet{wiklind1990}. 
\nb{The ages of these two components and their different metallicities lead to the conclusion that the period over which this galaxy sustained more or less constant SFR is probably much longer that what the 150\,Myr CST burst suggests. The galaxy core has seen substantial chemical enrichment throughout the last Gyr. }

The two older stellar populations have radially varying importance and the stellar light from the old population becomes dominant at radii larger than 30\arcsec{}. \citet{tikhonov2003} reached similar conclusion from their optical photometry studies.


\nb{In that respect, NGC404 is similar to the elliptical galaxy NGC205 where the central cluster, HubbleV, is younger and more metal-rich than the surrounding fields. There are also age and metallicity gradients for the field stellar populations of this galaxy \citep[see][for more details]{sharina2006}.}

\section{Discussion}
NGC404 has a very peculiar kinematical structure with two velocity gradient inversions (Figure \ref{rad_prof}) and the innermost region ($R<3$\arcsec{} \nb{or 45\,pc}) of the galaxy rotates in the same direction has the \hi{} gas \citep{delrio2004}. This inner core is also the region where the youngest stars are found (Figure \ref{sfh_rad}). Although \ulyss{} does not allow us to independently track the kinematics of each individual model component, it seems that this counter-rotation is a feature affecting only the newly-created stars. This suggests that the on-going star formation may have been triggered by the relatively recent ($\lesssim500$\, Myr) accretion of a nearby gas rich, low-mass companion. Our results are in perfect agreement with those of \citet{delrio2004} who argued that such an encounter must have happened approximately $0.5\times10^9$\,Gyr ago in the case of a catastrophic merger: this is required timescale to allow \hi{} gas to settle in the plane for the inner ring but not the outer one. This sort of encounter would result in star formation at the centre of the galaxy and exhibit the observed distinctive kinematical features.

Considering the scenario where dS0s are formed by the harassment and gas removal in late type spirals \citep{aguerri2005}, this places NGC404 in the awkward position of having \emph{acquired} gas rather than having it removed. However, since the galaxy obviously had multiple past encounters, it is not surprising to see a morphological transition occurring. The overall angular momentum of the old stellar population indicates that this galaxy was in rotation prior to its last merger. 
Furthermore, there is a significant, broadly distributed intermediate age stellar population (1.7\,Gyr) that could have been produced during a past, more or less extended period of active star formation (see Figure\,\ref{sfh_box}). 
We conclude that this system has been a spiral 1\,Gyr ago; the cold molecular gas found in the centre of NGC404 \citep{wiklind1990} is probably a relic from those earlier times \citep{temi2009}.

\citet[][in prep.]{sharina2010} found GCs in NGC404, and measured age and metallicity for one of them,
using medium-resolution spectra obtained with the SCORPIO spectrograph at the 6m telescope. It appears to be old (10\,Gyr) and metal-poor [Z/H]$\sim-1.7$\,dex.
Its properties are similar to those of the old stellar population of NGC404 and it is probably part of the entire system, having been formed during the initial star formation epoch.

\citet{maoz1998} classified NGC404 has a UV-bright Low Ionization Nuclear Emission line Region (LINER) galaxy and the emission line seen in Figure~\ref{example_fit} (namely \hb{}, \oiii{}, He\,{\sc ii} and N\,{\sc i}) are indeed typical of the high ionization range seen in narrow line Active Galaxy Nuclei (AGN). On the one hand, the apparent kinematic decoupling between the \oiii{} and \hb{} line emission (Figure \ref{rad_prof}) may be explained by this as the \oiii{} may have components coming from both the nebular component of the galaxy and its AGN; the width of the two lines would naturally decorrelate. On the other hand, since oxygen is mainly produced by very massive stars, the coincidence of the high \oiii{} activity with the GMC may be a sign of a massive star burst (maybe the formation of a new stellar cluster) and hence, a region of intensive stellar winds and greater gas turbulence. Interestingly, this region is near but somewhat offset from the centre of NGC404, reminiscent of circumnuclear rings often found in S0 galaxies \citep[e.g.,][]{silchenko2002}

The star formation in NGC404 is clearly confined within \nb{150\,pc} (10\arcsec{}) of its centre (Figure~\ref{sfh_rad}). Overall, it remains of low level: nowhere does the newly formed stars ($<1$\,Gyr) dominate the mass budget of the galaxy (it is less than 6\% in the centre). Unless future catastrophic events shape the evolution of this object, NGC404 should continue on this 'slow' evolutionary track and there is no reason to think that any significant star formation would occur at large radii. The mechanisms that lead this object from its past spiral morphology to its current dS0 state should continue their action until star formation definitely stops in this object, either because of fuel exhaustion or other gas removal mechanisms.

\section{Conclusion}
We have obtained spectroscopic observation along two position angles for the nearby dS0 galaxy NGC404. The spectra were analyzed using the \ulyss{} full spectral fitting package to find SSP equivalent ages, metallicities and kinematics for the different stellar populations.

Kinematically, the galaxy shows hints of a double radial velocity inversion a sign of multiple past encounters. This has probably triggered the evolution process leading the galaxy from a previous spiral morphology (1 Gyr  ago) to the current dS0 one. One of the latest merger event (with an \hi{} rich companion) would have triggered the most recent star formation episode which lasted for about 450\,Myr.

We resolved the star formation history of NGC404 into 4 events whose relative importances vary with radius. The youngest component is of high metallicity ([Fe/H] = $0.4\pm0.3$) and has ongoing star formation that lasted for the last 150\,Myr. Another young component of age 430\,Myr, has near solar metallicity ([Fe/H] = $0.1\pm0.1$). Both these component have negligible contribution to the spectra of NGC404 at a radius greater then 10\arcsec{} \nb{or 150\,pc}. The dominating stellar population is of intermediate age and near solar metallicity. It SSP equivalent parameters vary from 1.7\,Gyr and [Fe/H] = $-0.05\pm0.05$ in the centre to 2.6\,Gyr and [Fe/H] = $0.16\pm0.15$ \nb{at 370\,pc from the centre} (a radius of 25\arcsec{}). Finally, the last stellar population has an age of 12\,Gyr and [Fe/H] = $-1.26$. The decomposition of the history in a succession of bursts is an input characteristics of our model, and the time resolution is limited by the quality of the models and of the observations and by the intrinsic degeneracies of the problem. The metallicity  history was not constrained and reveal a realistic progressive enrichment. \nb{Considering the impact of internal dust absorption was shown to have negligible effect on the recovered star formation history and was therefore ignored to lower the number of free parameters in the fits. }

The kinematical and chemical history of NGC404 is consistent with it having evolved from a spiral morphology to a dS0 state. Future evolution will most probably continue this ongoing transition and, should we revisit this system in a few Gyr, we would most likely find a dE.

\begin{acknowledgements}
We are grateful to Fran\c{c}ois Simien who contributed to the early phases of this project, to Dmitry Makarov for his help during data reduction \nb{and to the anonymous referee for the useful comments}. AB acknowledges financial support from the South African Square Kilometre Array Project. \nb{MK acknowledges support by the Programa Nacional de Astronom\'{\i}a y
Astrof\'{\i}sica of the Spanish Ministry of Science and Innovation under grant \emph{AYA2007-67752-C03-01}.}
MS is partially supported by Russian Foundation for Basic Research grant \emph{08-02-00627}. We thank the Programme National Galaxies, from the French CNRS, for granting the observation time, and the staff from Observatoire de Haute-Provence for their support during the observations.

\end{acknowledgements}

\bibliographystyle{aa} 
\bibliography{404}   

\end{document}